\documentclass[a4paper,10pt]{article}
\usepackage[utf8x]{inputenc}
\usepackage[round]{natbib}
\usepackage{graphicx}
\usepackage{url} 
\usepackage[english]{babel}
\usepackage{placeins} 

\title{Integrating temporal and spatial scales: Human structural network motifs across age and region-of-interest size}
\author{
Christoph Echtermeyer\textsuperscript{1,2},
Cheol E. Han\textsuperscript{3},
Anna Rotarska-Jagiela\textsuperscript{4},\\
Harald Mohr\textsuperscript{5},
Peter J. Uhlhaas\textsuperscript{4},
Marcus Kaiser\textsuperscript{2,3,6,$\ast$}
}

\begin{document}

\maketitle

\noindent
\textsuperscript{1}Institute for Chemistry and Biology of the Marine Environment, Carl von Ossietzky University, Oldenburg, Germany \\
\textsuperscript{2}School of Computing Science, Claremont Tower, Newcastle University, Newcastle-upon-Tyne NE1 7RU, UK \\
\textsuperscript{3}Department of Brain and Cognitive Sciences, Seoul National University, Seoul 151-746, Republic of Korea \\
\textsuperscript{4}Department of Neurophysiology, Max-Planck Institute of Brain Research, Frankfurt a. M., Germany \\
\textsuperscript{5}Goethe University, Institute of Psychology, Frankfurt a. M., Germany \\
\textsuperscript{6}Institute of Neuroscience, The Medical School, Framlington Place, Newcastle University, Newcastle-upon-Tyne NE2 4HH, UK \\
\textsuperscript{$\ast$}E-mail: m.kaiser@newcastle.ac.uk
\\

\noindent
{\bf Running title:} Human structural connectivity patterns across time and space

\newpage
\begin{abstract} 
Human brain networks can be characterized at different temporal or spatial scales given by the age of the subject or the spatial resolution of the neuroimaging method. Integration of data across scales can only be successful if the combined networks show a similar architecture. 
One way to compare networks is to look at spatial features, based on fibre length, and topological features of individual nodes where outlier nodes form single node motifs whose frequency yields a fingerprint of the network. 
Here, we observe how characteristic single node motifs change over age (12--23 years) and network size (414, 813, and 1615~nodes) for diffusion tensor imaging (DTI) structural connectivity in healthy human subjects. 
First, we find the number and diversity of motifs in a network to be strongly correlated. 
Second, comparing different scales, the number and diversity of motifs varied across the temporal (subject age) and spatial (network resolution) scale: certain motifs might only occur at one spatial scale or for a certain age range. 
Third, regions of interest which show one motif at a lower resolution may show a range of motifs at a higher resolution which may or may not include the original motif at the lower resolution. 
Therefore, both the type and localisation of motifs differ for different spatial resolutions.  
Our results also indicate that spatial resolution has a higher effect on topological measures whereas spatial measures, based on fibre lengths, remain more comparable between resolutions.
Therefore, spatial resolution is crucial when comparing characteristic node fingerprints given by topological and spatial network features. 
As node motifs are based on topological and spatial properties of brain connectivity networks, these conclusions are also relevant to other studies using connectome analysis.

\end{abstract}

\\

\section{Introduction} 
The set of connections in the brain can be described as the connectome \citep{Sporns2005}. 
Connectome data is currently becoming available at different levels of the structural organization: from neuronal networks of connections between neurons to fibre tract networks between brain regions.
These data promise to give valuable new insights, but analyses integrating different data sets are challenging.
One problem of data integration across labs is that the raw data might show different voxel resolutions due to different magnetic field strengths of the MRI (magnetic resonance imaging) scanner.
In addition, different labs might use different parcellation routines or brain atlases. 
This leads to different spatial resolutions and consequently to different structural connectivity networks (eventually with different numbers of nodes) \citep{Zalesky_2010b, Bassett2010, Fornito2010, Hayasaka2010}.
Another problem of data integration is caused by age differences between subjects.
There is a certain evidence that characteristics of the connectome vary over age \citep{Fan2010, Hagmann2010b, Fair2009, Uhlhaas2009}.
Given a network analysis result of two studies, say concerning small-world features \citep{Watts1998}, are differences due to the study population's differences in connectivity or are they due to differences in age or network resolution? 
In this article, we observe how topological and spatial network features change across age and network resolution for structural connectivity, based on diffusion tensor imaging (DTI), in healthy human subjects.
Using single node motifs, we also apply a network mapping technique, which yields a compact and easily comparable representation of complex network structures.

\subsection{Scales of a hierarchical organisation}
Hierarchy is a central feature in the organisation of complex biological systems and particularly the structure and function of neural networks \citep{Kaiser2010editorial, Kaiser2010}.
In neural networks, the term hierarchy may be understood in several different ways and can apply to topological, spatial, temporal as well as functional properties.
These features can be observed at different scales or resolutions.

\subsubsection*{Temporal scales}
Network analysis is often presented as the analysis of a network at a given time, but many networks change both over time and with temporal resolution.
Structural connectivity, given by anatomical connections between neurons or brain regions, may change due to activity-dependent plasticity \citep{Butz2009a}.
Functional connectivity, given by correlations in the activity patterns of network nodes, can change for different tasks for a subject or due to different system states (e.g. awake vs. asleep). 
In our case, connectivity might also change over longer time-scales, as part of brain development.

Recovered networks can appear differently depending on temporal resolution.
At a fine-grained temporal resolution of a few hundred milliseconds, cortical tissue can show distinct oscillations, e.g. in the $\alpha$, $\gamma$, or $\delta$~frequency bands \citep{Roopun2008}.
For more coarse-grained time-resolutions with sample lengths of several seconds or minutes such distinct frequency peaks disappear and the frequency distribution displays a $1/f$~power-law behaviour instead \citep{Buzsaki2006}. 

Here, we observe temporal scale in terms of snapshots of structural connectivity at different points in time.
The human brain undergoes large structural developments until the age of around 20--25~years.
These developments coincide with large functional changes during child development and during the teenager years.
They further coincide with changes in functional connectivity \citep{Uhlhaas2009}. 
Such changes during brain network development are not only of interest for the study of healthy subjects, but also for many psychiatric disorders like schizophrenia \citep{Uhlhaas2010a}, which have an onset around the time when the brain network matures (age 18--25 years).
We therefore test the potential of single node motifs to integrate information across temporal network 'snapshots'.

\subsubsection*{Spatial scales}
The network structure not only depends on the time of a 'snapshot' but also on the spatial resolution.
One can distinguish the micro- and macro-connectome as the connectivity between individual neurons and connections between brain regions, respectively \citep{DeFelipe2010}.
In this study of structural connectivity we focus on the macro-connectome level based on MRI data (see \citet{Seung2009} for more information about neuronal connectivity instead). 

The macro-connectome can be analysed at different levels of spatial resolution and thus different levels of brain network organisation.
With a low spatial resolution, it corresponds to connectivity between brain areas where network nodes correspond to brain regions.
The resulting networks show features of modular and small-world networks  \citep{Hilgetag2000a,Sporns2000a} and typically consist of up to 100 nodes per cortical hemisphere for the primate brain.
At the mesoscale, connectivity between regions of interest of the same size, e.g. 1cm$^2$ cortical surface area, can be studied.
Such networks consist of around 1,000~nodes for the human brain \citep{Hagmann2008}.
With even higher resolution, the microscale of the macro-connectome could be studied, i.e. connections between cortical columns \citep{Mountcastle1997}.
However, such networks with nodes representing one cortical column each would consist of 1,250,000~nodes for the human brain (both hemispheres, not including subcortical structures) and identifying columns with diameters ranging from 200$\mu m$ to 1$mm$ \citep{Hubel1977,Horton2005} in humans is (currently) beyond the reach of standard MRI. Such extremely high resolutions might only be achieved with higher magnetic fields or extremely high magnetic fields in {\em post mortem} studies.

We therefore study the global and regional scale of the macro-connectome.
Importantly, this level allows us to assess whether motifs which are characteristic for a certain cortical region are also characteristic for all parts of that cortical area or whether a region displays a variety of motifs at higher spatial resolutions.
Also, changes in spatial resolution affect spatial measurements, which form part of the motif detection.
It is often interesting to observe how many connections go to nearby targets and how many extend over a long distance, potentially linking different components of the neural network.
This can be readily observed using a histogram of the connection lengths of a network.
It is known that the probability that two neurons are connected decays almost exponentially with distance \citep{Hellwig2000, Kaiser2009, Schuez2005}: connections over a long-distance are less likely than short-distance connections.
In different organisms, ranging from neuronal connectivity in {\em C.~elegans} and layers in the rat visual cortex to fibre-tract connectivity in the macaque, the actual distribution was approximated best by a Gamma probability density function \citep{Kaiser2009}.
Looking at different spatial resolutions within humans, we would expect that the proportion of short-distance connections relative to long-distance connections increases when the spatial resolution is increased.
We will therefore observe how the length distributions change over different spatial scales.

\subsection{Single node motifs}
Networks can be analysed following different concepts:
To investigate a single network, for example, analysis can focus on its nodes, comparing them through {\em connectional fingerprints} based on the absolute values of node properties (e.g. node degree or local clustering coefficient) \citep{Passingham2002}.
On a more abstract level, nodes of brain networks can be regarded with respect to their different organisation levels.
These approaches look at one network of a particular subject, but networks can also be compared across different subjects---or different networks might exist for the same subject across time.
However, mapping between networks becomes difficult if the number of nodes and edges varies, such that direct comparisons (i.e. link-to-link or node-to-node) between structures are ruled out.
In contrast to such methods of cross-comparison, the approach used in this paper is less sensitive to these changes: 
Although a network with a slightly higher number of nodes could show additional motifs, most nodes are non-outliers making such changes less likely. 
Similarly, a network with slightly more edges might show absolute changes in network features without affecting the outliers of the feature distribution leading to similar single node motifs.
Again, these comparisons of the number and classes of motifs become possible for non-identical numbers of nodes or edges but only if the numbers do not differ too much.
These motif-classes are significantly different from each other and they allow to classify network nodes at the same level even if they appear in different networks.

The concept of {\em node-motifs}---a combination of local network features--- complements that of {\em network-motifs}, which are specific connectivity patterns that have been used to characterise networks before \citep{Milo2002}.
An example for a node motif are highly connected nodes or hubs that affect spreading phenomena and can be important components of a network.
More complex, multi-dimensional node-motifs, which are characterised by multiple features in combination, specify nodes more comprehensively.
With these more precise descriptions, new kinds of motifs can be formulated and \citet{Costa_2009} presented a routine for their detection and specification.
We proposed first improvements and procedures for automating parameter choices for this method \citep{Echtermeyer2011}, such that large numbers of networks can be processed and compared to each other.
This enhanced technique is applied to brain networks in this paper.

\subsection{Comparing motif counts across networks}
The usefulness of mapping brain networks (which itself are abstract representations) to motif-classes is demonstrated for two scenarios:
First, regional connectomes of 53 children (aged 12--23 years) are compared to show how the fingerprint of a DTI network can change over time.
Second, we test the effect of different connectome-resolutions on the resulting network-fingerprint.
Therefore, DTI networks at different levels of organization were generated by changing the level of granularity from 414 regions of interest~(ROIs) to up to 1,615~ROIs.

Comparing networks from the same subject but at different granularities has so far been complicated by the critical dependence of standard approaches on the resolution of the network \citep{Antiqueira2010}.
We show how our novel node fingerprints can be used to compare different levels of connectome organization.

Note that comparing motifs has several advantages over comparing network features.
Features, such as the small-world measures of clustering coefficient and characteristic path length, critically depend on the edge density of a network.
Therefore, changes in these measures could be either due to changes in network organization (e.g. modularity) or due to a difference in the number of connections.
To identify the changes related to network organization, the edge density needs to be constant when comparing networks.
However, this leads to a severe problem: when transforming raw data of connection strengths between nodes to binary data, each network needs a different threshold for establishing a connection \citep{Rubinov2010}. 
To avoid the problem of either having different edge densities or different thresholds, researchers increasingly use weighted networks without thresholding \citep{Bullmore2011}. 
For our analysis, we expect that differences in edge density are not only due to inter-subject variability, as in other studies, but are more heavily influenced by changes across different temporal or spatial scales. 
Therefore, we observe how network measures, and as a result node motifs, change over scales.
These changes can be due to changes in organisation or edge density.

\begin{figure}[h]
 \centering
 \includegraphics[width=0.95\textwidth]{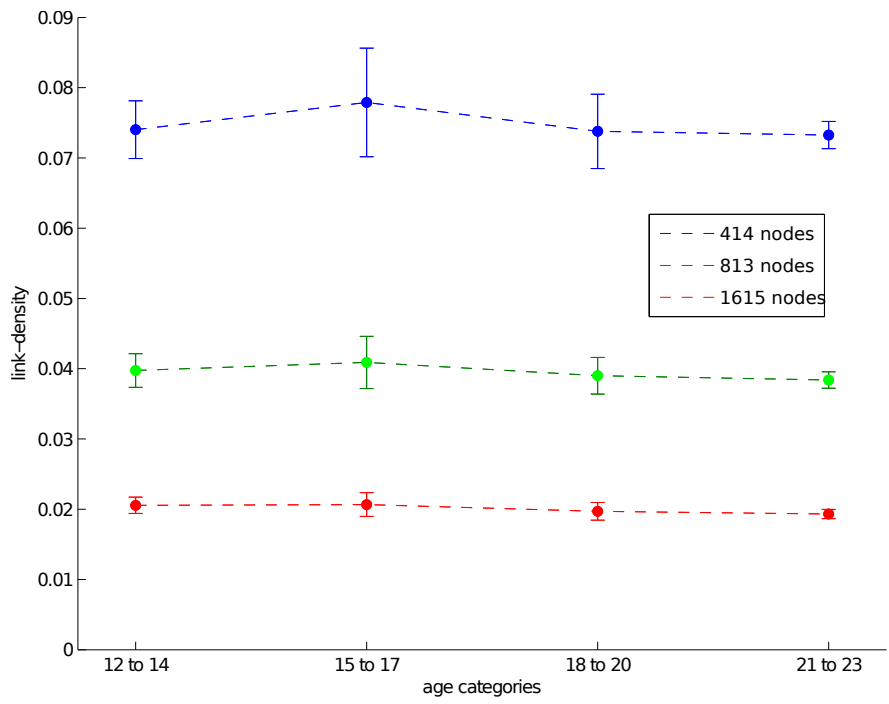}
 \caption{
   Edge density for different spatial resolutions, given by the number of network nodes, and different age groups.
   Data points show the average edge density of all networks belonging to a certain age group and spatial resolution.
   Corresponding standard deviations are shown by error-bars.
 }
 \label{fig_edge_density}
\end{figure}

Although a detailed analysis is outside the scope of this paper, we show the relation between both scales in Figure~\ref{fig_edge_density}.
%
%
Whereas edge density decreases with network size, there are no significant changes between age groups when looking at networks with the same spatial resolution. 
However, we would like to point out that (a)~motifs change over time and can therefore be used to find characteristic changes in the developing brain and (b)~motifs change over spatial scales, therefore comparisons between studies are only meaningful when the same (or a comparable) spatial resolution has been used.

\section{Materials and Methods} 
\subsection{Recruiting and Data Recording}
We recruited 53 participants (age between 12 and~23~years) from local community and high schools.
The study received ethics committee approval and written informed consents were obtained from the participants' parents for under-age participants, or from themselves if older than 18 years after study procedures were described.
Participants were screened for a history of psychiatric and neurological disorders and current drug abuse using ophthalmological assessment including monocular and binocular visual acuity.
Participants were grouped into four age-categories (Table~\ref{table_age_cat}). 


A 3~Tesla scanner (Siemens Trio, Erlangen, Germany) was used at the Brain Imaging Centre (BIC), Frankfurt am Main, Germany.
We recorded T1 weighted MRI (voxel-size 1x1x1mm\textsuperscript{3}) and Diffusion Tensor MRI (voxel-size 2x2x2mm\textsuperscript{3}).
To minimize head motion, we used tightly padded clamps on the eight-channel head coil.
The T1 weighted MRI was recorded with the following parameters: 176 slices, Field of View~(FoV) 256mm, repetition time~(TR) 2250ms, and echo time~(TE) 2.6ms.
Three Diffusion Tensor Images per participant were recorded with the following parameters: 60~slices, FoV~192mm, TR~7600ms, TE~91ms, 60~directions b-vectors with b-factor of 1000 and 10 b0~images.

\subsection{Data Processing and Network Construction}
We used freesurfer to obtain surface meshes of the boundary between grey matter and white matter from T1 anatomical brain images (\url{http://surfer.nmr.mgh.harvard.edu}), and Diffusion toolkit along with TrackVis \citep{WANG_2007} to obtain stream-line tractography from diffusion tensor images.
We registered surface meshes into brains of the diffusion tensor images to extract networks. 
Freesurfer provides parcellation of anatomical regions of cortices (33 for each hemisphere) based on the Deskian atlas \citep{Deskian_2006, Fischl_2004a} and subcortices \citep{Fischl_2002,Fischl_2004b}.
We subdivided those anatomical regions of cortices into many of similar surface area using our own algorithm.
It is known that the size of each region of interest~(ROI) affects network connectivity and various network measures \citep{Zalesky_2010b, Hagmann_2007}.
The core of our algorithm is an expectation-maximization~(EM) algorithm.
First, we choose the expected number of subdivisions (400, 800, and 1600) for cortical regions (we do not subdivide subcortical areas) and decided the number of subdivisions based on the ratio of each surface area over whole surface area in the average subject.
Then we sub-divide each anatomical cortical region to the assigned number of subdivisions using the EM~algorithm, minimizing variance of subdivisions' surface area sizes. 
The ratio of variance to mean does not exceed 10\% in the average subject. 
In each subject, this ratio was higher (400:  414.78mm\textsuperscript{2}~$\pm$~140.06, 33.77\%; 799: 207.65mm\textsuperscript{2}~$\pm$~77.51, 37.33\%; 1601: 103.63mm\textsuperscript{2}~$\pm$~44.33, 42.78\%, statistics of cortical ROIs only) because the size of each subject's anatomical region varied (2468.06mm\textsuperscript{2}~$\pm$~1813.10mm\textsuperscript{2}, 73.46\%, which represents individual subjects' regions were 12\%~$\pm$~30\% bigger than the average subject's; ratio of variance to mean is 26.49\%). 
Another useful statistic is the ratio of interquartile range to median: larger values represent higher dispersion. 
The values (Table \ref{table_surf_area_stat}) are smaller than those determined by \citet{Fornito2010}.

\begin{figure}[h]
 \centering
 \includegraphics[width=0.9\textwidth]{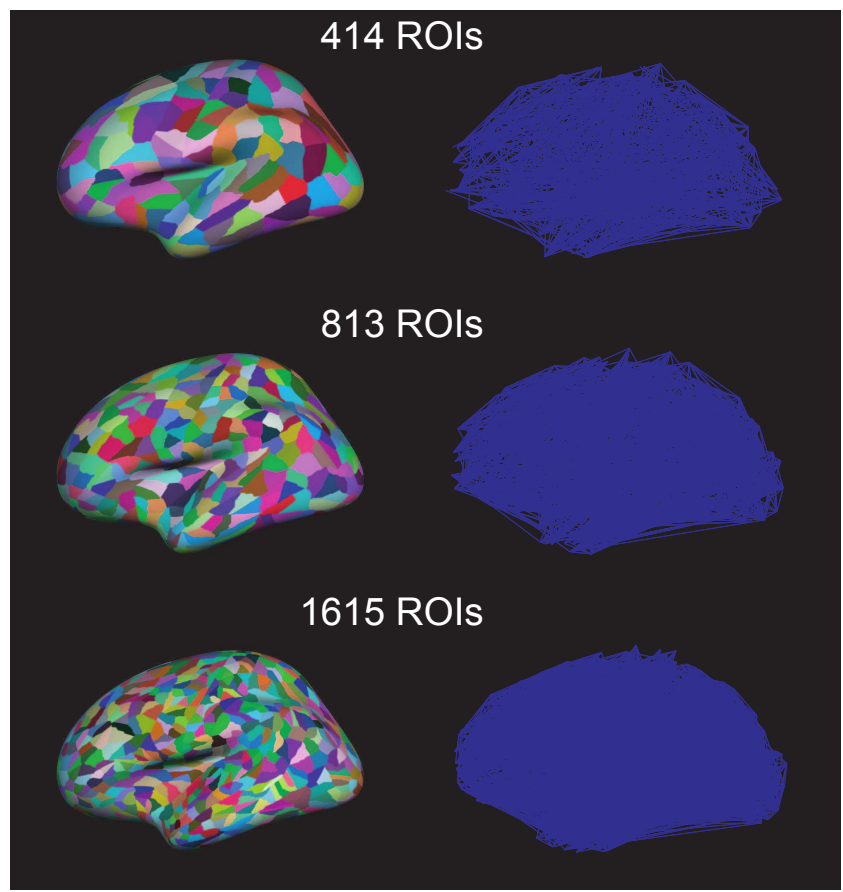}
 \caption{
   Region of Interests (ROIs) on the brain (left) and the resulting networks (right) with different numbers of ROIs.
 }
 \label{fig_left_hem}
\end{figure}

We projected this new parcellation template into each subject's surface.
Using this projection procedure we kept topological consistency across all the subjects.
We also included a few selected subcortical areas in the ROI list: Nucleus accumbens, Amygdala, Caudate, Hippocampus, Pallidum, Putamen, and Thalamus.
Thus the actual number of ROIs (414, 813, and 1615, respectively) is slightly larger than specified.
These three numbers of ROIs on the brain and resulting networks are shown in Fig.~\ref{fig_left_hem}.
%
%
Before further processing the diffusion tensor images, we serialised three diffusion tensor images and b-vectors for each participant; thus, a collated image of each subject has 30 b0-images and 180 diffusion images. 
The eddy current was corrected through FSL (\url{http://www.fmrib.ox.ac.uk/fsl/}).
We used Diffusion toolkit along with TrackVis \citep{WANG_2007} with Fiber Assignment by Continuous Tracking (FACT) algorithm \citep{Mori_1999} and 35 degree of angle threshold.
After registering the surface meshes (FreeSurfer, \url{http://surfer.nmr.mgh.harvard.edu/}) of each subject to his/her DTI, we generated cortical volume ROIs, which are voxels in the gray matter.
Also the selected subcortical areas (see above) were registered to DTIs and used as ROIs.
Then, we used the UMCP toolbox (\url{http://ccn.ucla.edu/wiki/index.php/UCLA_Multimodal_Connectivity_Package}) to obtain connectivity matrices from the defined and registered ROIs and tractography.
This counts the number of fibres from a tractography from the Diffusion toolkit, between all pairs of defined ROIs' grey matter; the fibre-counts between all ROI-pairs yield the weight-matrix.
We also computed the average fibre lengths between ROIs (if there is no connection between a pair, the length is set to zero). 
The fibre length is based on the actual three-dimensional trajectory of the fibre tract and may be larger than the Euclidean distance between connected nodes. Note that long fibre length usually also corresponds to a large Euclidean distance between ROIs: earlier studies in fibre tracts in primates have shown that 85\% of the fibres go in a straight line or are only slightly curved \citep{Hilgetag2006}. However, only a fully bended (U-turn) fibre could result in a long fibre tract connection to spatially nearby nodes.  

DTI fibre-tracking yielded weighted matrices where weights indicate the probability of connections between corresponding ROIs.
Non-zero weights have been interpreted as an indication for a link when constructing the brain networks.
The resulting adjacency matrices were similarly sparse for each network-resolution (Fig.~\ref{fig_edge_density}, average link-densities 7.5\%, 4.0\%, and 2.0\% for networks of 414, 813, and 1615~nodes, respectively).

\subsection{Network Analysis}
Single node-motifs were identified from the constructed networks by applying a technique presented by \citet{Costa_2009} with additions in order to choose parameters automatically \citep{Echtermeyer2011}.
The method yields a global characterisation of the network based on its local properties.
Here, each node is quantified by 9~different local measures, which are motivated in the next paragraph.
We chose such a large number of local measures, some of which share similarities, to ensure that as many network features as possible are considered. 
Redundancies, where two features are highly correlated, are removed during the node-motif detection, which is explained further below.

A nodes' {\em degree} is its number of connections to other nodes; dividing this by the number of all links in the network yields the {\em normalised node degree}~$K$.
The average over all neighbours' degrees is called the {\em normalised average degree}~$r$.
(Nodes that are directly linked to each other are called {\em neighbours}.)
Degrees of a node's immediate neighbours can further be taken into account by their respective {\em coefficient of variation}~$cv$.
Connectivity among neighbours of a node is quantified by the {\em clustering coefficient}~$cc$, which reflects how many of all possible connections between neighbours actually exist~\citep{Watts1998, Kaiser2007njp}.
The {\em hierarchical clustering coefficient} of level two~$cc_2$ extends this concept to connections between neighbours' neighbours~\citep{Costa2006}.
To what degree a node's neighbours connect to the same target is quantified by the {\em locality index}~$loc$, which is based on the matching index \citep[e.g.][]{Kaiser2004}.
The 6~measures mentioned so far quantify topological aspects of the network, while the next 3~measures also take into account spatial features.
The {\em average connection length}~$acl$ for a node is the average length of fibre tracts to its neighbours and the furthest away neighbour determines the {\em maximum connection length}~$mcl$.
The average over connection lengths to nodes two steps away, i.e. neighbours' neighbours, is called the {\em average indirect reach}~$air$.
Further details and measures can be found in the literature~\citep{Albert2002, Newman2003, Newman2006, Costa2007a}.

\begin{figure}[h]
 \centering
 \includegraphics[width=\textwidth]{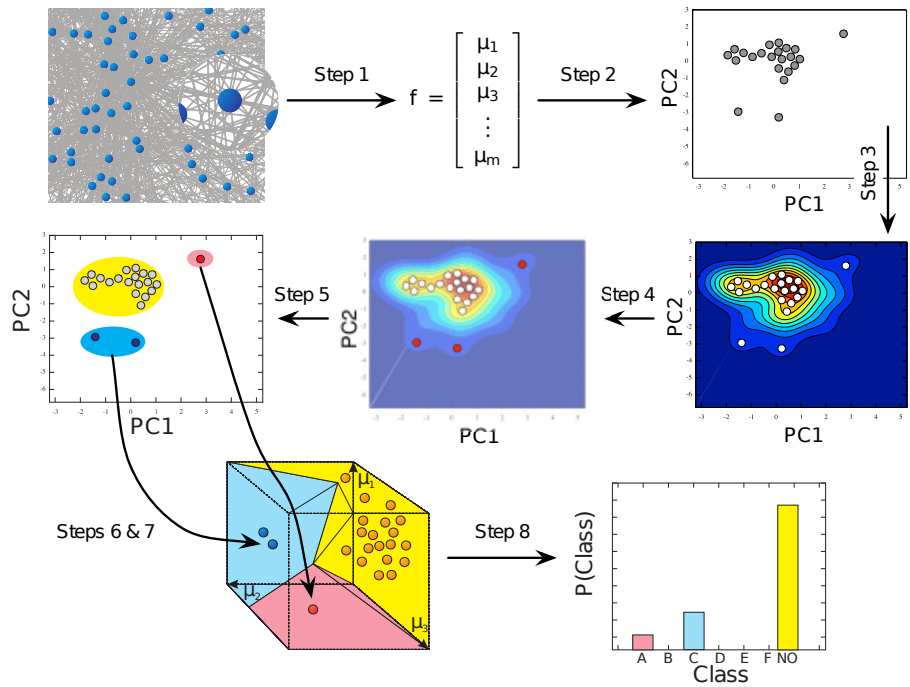}
 \caption{
  Figure from \citet{Echtermeyer2011}.
  Illustration of network-analysis work-flow:
  High dimensional characterisation of all network nodes through local network measures~$\mu_i$ (Step~1) is compacted to two dimensions (Step~2) in order to estimate a probability distribution (Step~3), which is used to identify nodes with uncommon features (Step~4).
  All nodes are grouped (Step~5) to form high dimensional motif-regions (Step~6), which are eventually joined, if too close to each other (Step~7).
  The number of nodes in each motif-region yields a fingerprint of the network (Step~8).
 }
 \label{fig_workflow}
\end{figure}

The motif-detection workflow starts by applying the above listed network measures (Fig.~\ref{fig_workflow}):
%
%
Local network measures are calculated for all nodes of the network, which yields a 9-dimensional characterisation ({\em feature vector}) for each node.
This high-dimensional representation is simplified by exploiting similarities between different measures \citep{Costa2007a}:
Using principal component analysis~(PCA)~\citep[Chapter~8]{Johnson2007}, correlation between measures is removed and the feature vectors are reduced to two dimensions~({\em PCA-plane}).
The position on the PCA-plane characterises the nodes and allows for comparisons:
Nodes close to each other share similar features whereas well separated ones differ.
The next step is to estimate how likely specific combinations of features are.
This is done by smoothing over points in the PCA-plane using the Parzen window approach~\citep[Chapter~4.3]{Parzen1962, Duda2001} with width of the smoothing kernel scaled according to the standard deviation along the corresponding principal component axis~\citep{Echtermeyer2011}.
Thereby, each node is assigned a probability, which is used to distinguish those with common and rare features.
The $w$~many nodes with uncommon properties, termed {\em outlier nodes}, were determined as follows.
Given all nodes' probabilities~$ p = (p_k)_{k=1,\ldots,n} $ (sorted
increasingly), the respective mean~$\bar{p}$, and standard deviation~$\sigma(p)$,
the number of singular nodes~$w$ is chosen as
\begin{equation} \label{eq_w}
    w = \arg\max_{k \ : \ p_k < \ \bar{p} - \sigma(p) }{ p_{k+1} - p_k } \ ,
\end{equation}
or $ w = 0 $, if all probabilities are greater than~$\bar{p} - \sigma(p)$.
Nodes corresponding to the smallest $w$~probabilities are the identified outliers.
Next, the outlier nodes were assigned to clusters each of which represents a different {\em motif-group}, i.e. a set of features.
This was done by first centring equally sized ellipses on each point and
determining their overlap.
Maximal sets of nodes whose ellipses form a connected area yield both the
number~$k$ of motif-groups as well as the groups themselves.
(The above mentioned mechanisms have been evaluated and discussed in detail in an
earlier publication \citep{Echtermeyer2011}.)
Motif-groups correspond to 9-dimensional {\em motif-regions} in feature space, such that all nodes within a motif-group are similar with respect to the local network measures used in the beginning.
In the final step, motif regions from all networks were reduced to the final 9 motif regions by subsequently joining the closest two of them.
The pairwise distances have been determined with a modified Mahalanobis measure~\citep{Mahalanobis1936}.
The number of nodes that express a particular motif yields a {\em fingerprint} of the network, which characterises the network and allows for comparisons \citep{Costa_2009}, e.g. by considering the {\em motif-diversity}, which is the number of unique motifs~$k$ that were identified.
Another possibility is to compare the number of outlier nodes~$w$.

\section{Results} 
\subsection{Motif Changes with Age}
\begin{figure}[h]
 \centering
 \includegraphics[width=\textwidth]{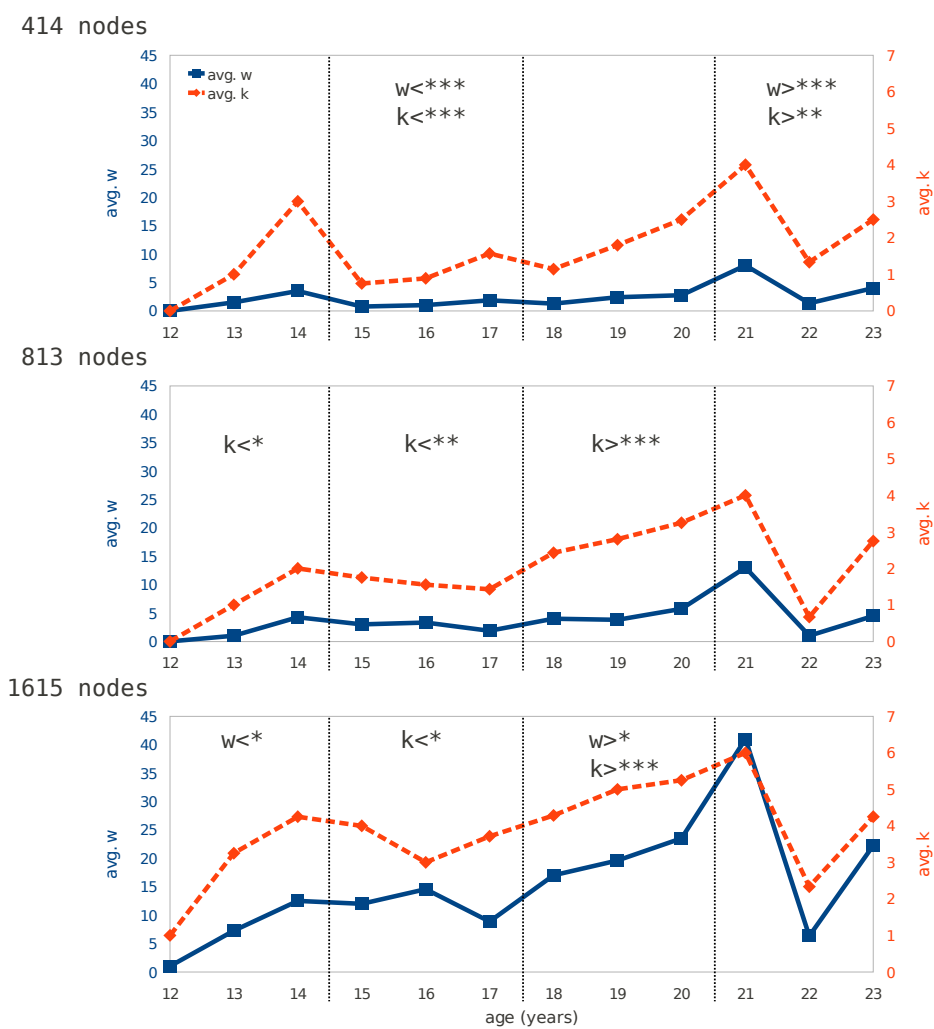}
 \caption{
   Motif-expression changing with age:
  Both the number of outlier nodes in a network~$w$ (solid blue line) and the diversity of motifs expressed~$k$ (dashed orange line) vary with subject age.
   The time-dependent patterns of~$w$ and~$k$ are shown for different network resolutions (rows).
  Age-groups indicated by dashed vertical lines.
  Significantly de- or increased values for~$w$ and~$k$ are indicated by symbols~$<$ and~$>$, respectively (*~90\%, **~95\%, and ***~99\% significance). 
  Note that our data only include one subject aged 21 and further data would be needed to confirm significance of deviations at this age.
 }
 \label{fig_w_k_distribution}
\end{figure}

Networks were first analysed with respect to age-dependent changes.
Therefore, outlier motifs were determined for all networks to see whether certain motifs might only be expressed in subjects of similar age.
This was not the case, but the total number of outlier-nodes~$w$ shows a characteristic pattern with age (Fig.~\ref{fig_w_k_distribution}, solid blue lines).
%
%
Subjects of young ages (13--14~years) show 3--4 outlier nodes on average---a number that is reduced throughout the teenager years and thereafter (15--20~years) until a sudden and predominant peak at the age of~21.
Subjects aged 22--23~years have again fewer, but still several outlier nodes.
This pattern is qualitatively preserved for different network resolutions (Fig.~\ref{fig_w_k_distribution}~top to bottom).
Quantitatively, the number of outlier nodes~$w$ rises with larger networks.

To determine which changes are statistically significant a random permutation test has been applied: Subjects were randomly assigned an age (12--23~years) while ensuring that each age-group had the same size as for the original data.
Subsequently, each group's mean~$w$ and~$k$ were compared to those of the original data to compute the P-values (using a total of 500,000~permutations).
Results, shown directly in Fig.~\ref{fig_w_k_distribution}, represent how far each group is to the average over age.
Note that except for the lowest network resolution, the large spike at age~21 does not lead to significant increase in~$w$ (and~$k$).
This is because our data only include one subject aged~21 and further data would be needed to confirm that the deviations at this age are indeed significant across all scales of resolution.

A pattern that is very similar to that of the outlier number~$w$ is the diversity~$k$ of expressed motifs.
The number~$k$ is high and low for nearly the same ages as was the number of outliers~$w$ before (Fig.~\ref{fig_w_k_distribution}, dashed orange lines), which is also reflected by the strong correlation between~$w$ and~$k$ (Pearson correlation 0.94, 0.88, and 0.87 for 414-, 813-, and 1615-node networks, respectively).
Peaks at low (13--14~years) and high (21~years) ages can be seen with a dip in between.
This pattern fits networks with 414~nodes best, whereas networks with the higher resolution (813 and 1615~nodes) show an increasingly flat profile except for the later peak (21~years).
Note that absolute motif-diversity~$k$ is relatively similar for all network resolutions, which was not the case for the actual number of outliers~$w$.

Both, number of outliers~$w$ and motif-diversity~$k$ are interesting on their own, but their strong correlation provides additional information.
Although the absolute values differ, similarity of the curves' profiles is independent of network resolution.
We thus find the very robust effect that motif-diversity is coupled to the number of outlier nodes.
In other words, whenever more outliers occur, they also become more diverse.

\subsection{Motif Changes with Network Resolution}
The previous section has already shown that network resolution affects outlier quantity, $w$, but not the relative distribution across the age-range (Fig.~\ref{fig_w_k_distribution}, solid blue lines).
Different from that, motif-diversity---i.e. how many different motifs are expressed, $k$---is nearly invariant on an absolute scale, in contrast to its relative distribution (Fig.~\ref{fig_w_k_distribution}, dashed orange lines).
We tested whether the number~$w$ and diversity~$k$ of outliers, at different spatial scales, are correlated with cognitive skills of our subjects.
Unfortunately, measures of intelligence yielded by four sub-scores of the Wechsler intelligence test were only available for 44 out of our 53 subjects.
For all comparisons, the Pearson correlation was $r \leq 0.5$.
Therefore, we cannot confirm a link between intelligence and~$w$ or~$k$.
However, intelligence might still be related to a specific combination of motifs; a possibility which could only be tested given a larger cohort of subjects.
Next, we analyse the actual frequencies with which motifs are expressed by different age groups (Table~\ref{table_age_cat}).

\begin{figure}[h]
 \centering
 \includegraphics[width=0.6\textwidth]{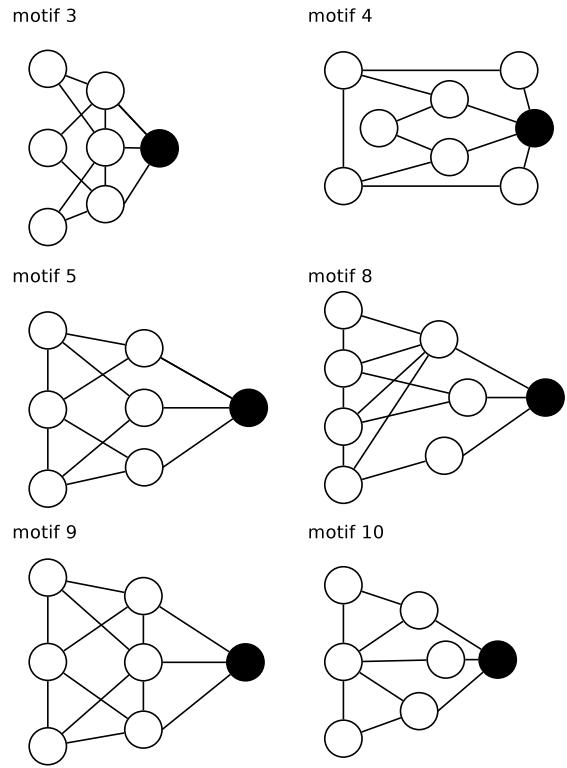}
 \caption{
   Stylised illustration of most frequent motifs (Table~\ref{table_motif_properties}):
  \textbf{motif~3}~{\em Provincial node} with connections in the direct (high~$cc$) but not the larger neighbourhood (low~$cc_2$, $air$, and $mcl$).
  \textbf{motif~4}~{\em Provincial hub} with more connections than its neighbours (high~$r$) that are less connected between themselves (low~$cc$).
  \textbf{motif~5}~{\em Global bottleneck} with few (low~$K$) but long-range connections (high~$acl$, $mcl$, $air$) that reach beyond the local neighbourhood (low~$cc$). 
  \textbf{motif~8}~{\em Global uniform bottleneck} sharing properties of motif 5 but connected to nodes with similar degrees (low~$cv$). 
  \textbf{motif~9}~{\em Global local bottleneck} sharing properties of motif 5 but also having well-connected neighbours (high~$cc$) thus better informing local circuits.  
  \textbf{motif~10}~{\em Provincial bottleneck} with only short-range connections (low~$mcl$) and few connections between its neighbours (low~$cc$) leading to a large influence on the local circuit.
 }
 \label{fig_motif_sketch}
\end{figure}

Using the same raw DTI-data, networks have been constructed with different resolutions, i.e. differently sized ROIs, for each subject.
More fine grained ROIs parcellate the brain into more elements, each of which is represented by a network node.
We were interested in the effect that different network resolutions have on the occurring node-motifs (Table~\ref{table_motif_properties}, Fig.~\ref{fig_motif_sketch}) and therefore, networks with 414, 813, and 1615~nodes were analysed and compared to each other.



\begin{figure}[h]
 \centering
 \includegraphics[width=0.95\textwidth]{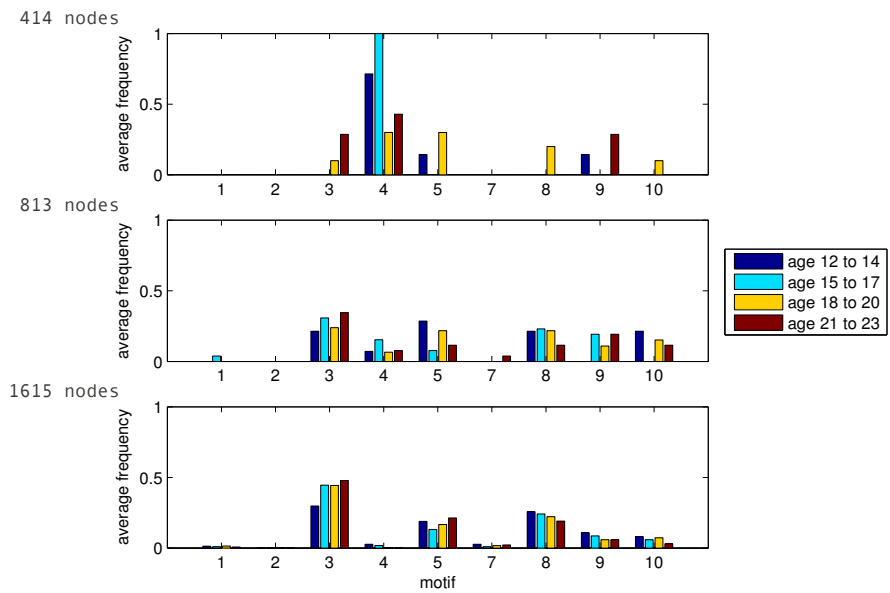}
 \caption{
   Motif-expression changing with network-resolution.
   Plots show distribution of outlier nodes among motifs~1--5 and 7--10.
   Motif~6 (not shown) corresponds to the remaining 98\% network nodes with common features (regular nodes).
 }
 \label{fig_dep_net_res}
\end{figure}

The results show that motif-expression strongly depends on network-resolution (Fig.~\ref{fig_dep_net_res}).
Whereas motif~4 is most frequent for the 414-node resolution, it is nearly absent in the highest resolution of 1615~nodes.
Other motifs also show large fluctuations between spatial resolutions with either increasing or decreasing frequency.
Motifs can also be characteristic for a single scale.
For example, motifs~8, 9, and~10 are most prevalent across age groups at an intermediate resolution of 813~nodes.
This shows that the spatial resolution is critical for assessing the frequency of motifs and, presumably, also for its underlying spatial and topological features.

Motif counts also depend on age. 
For 414~node networks, many motifs only occur for one or two out of the four age groups.
Whereas this might be influenced by the low number of nodes in the network, we also see age-based motif frequency changes for higher spatial resolutions.
Looking at the resolution of 1615 nodes, the number of nodes showing motifs~3 and~5 increases with age whereas motifs~8 and~9 become less frequent.
These age-dependent trends are less visible at 813-node resolution and they disappear for the 414-node networks.


\subsection{Fibre length distribution for different network sizes}
\begin{figure}[h]
 \centering
 \includegraphics[width=0.95\textwidth]{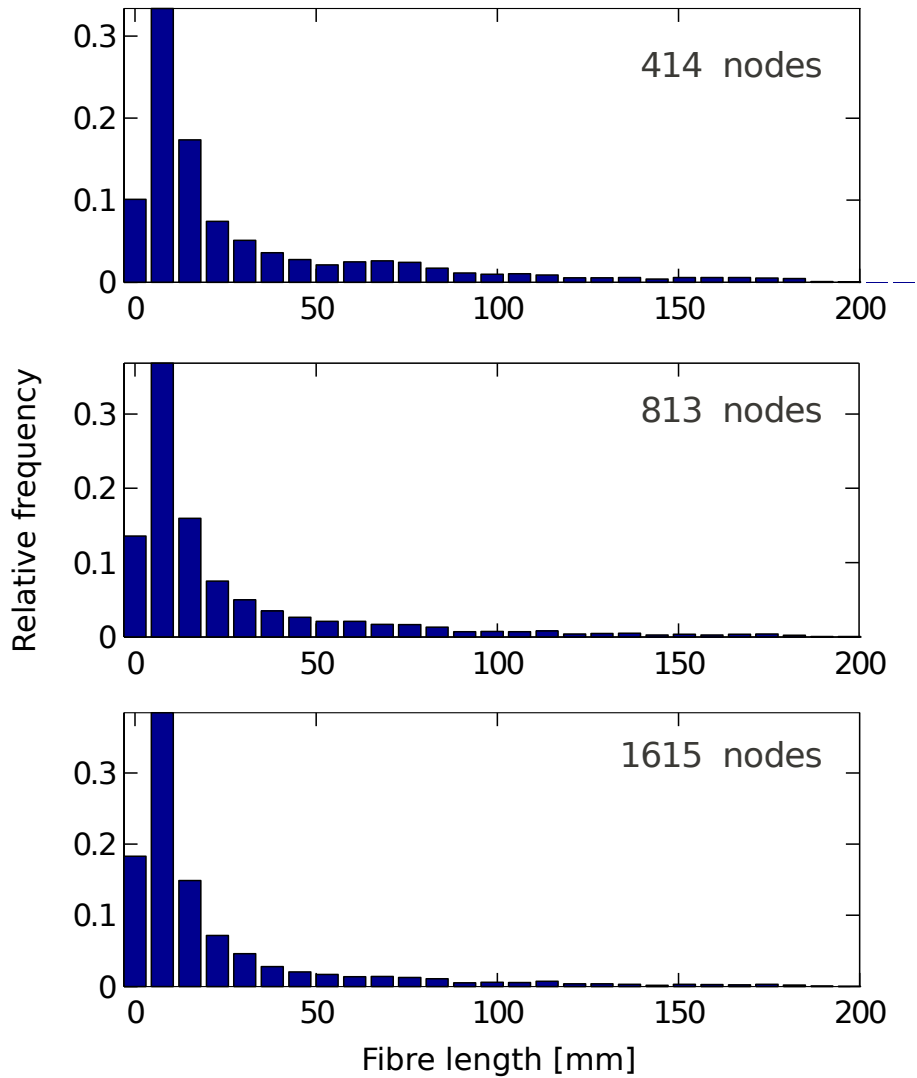}
 \caption{
  Fibre length distribution, as the length of the trajectory in mm, for different network resolutions. 
  Relative frequencies for low (414 nodes, top), medium (813 nodes, middle), and high (1615 nodes, bottom) spatial resolution. 
  Note that longer fibres ($>$200mm) occurred so infrequently that corresponding bars (not shown) would be invisible.
  }
 \label{fibre_length_dist}
\end{figure}

A possible explanation for the observed dependence on network-resolution could be fibre-length based network measures (namely: average connection length, maximum connection length, and average indirect reach).
These measures take into account the determined fibre trajectory between ROIs, i.e. their fibre tract length.
It is thus to be expected that the number of ROIs impacts on the corresponding fibre lengths.
We tested this by observing the fibre length distribution for different network resolutions (Fig.~\ref{fibre_length_dist}).
Across all resolutions, short fibres are more frequent than long fibres. 
This exponential tail, which can be fitted by a Gamma distribution \citep{Kaiser2009}, has been reported before not only at the level of fibre tracts \citep{Kaiser2004c}, but also at the level of connections between neurons within cortical areas \citep{Hellwig2000}.
One notable difference between spatial resolutions is the increase in the number of short fibres (length$<$10mm) increasing by more than 50\% from the lowest to the highest spatial resolution.
This can be due to the smaller size of network nodes where surface regions that before belonged to the same node are now separate nodes on the surface. 
Therefore, a short fibre tract between these nodes becomes feasible.
Looking at the maximum range of fibres, spatial resolution does not seem to influence maximum fibre lengths.
However, differences in average fibre length, in particular due to an increased number of short fibre connections, might occur for particular nodes.    
The pattern for average indirect reach is more complex and can be influenced by changes in both short and long fibres.
Overall, given a three-fold increase in spatial resolution, the fibre length distributions remain remarkably similar. 


\subsection{Motif consistency across spatial scales}
\begin{figure}
 \centering
 \includegraphics[width=0.95\textwidth]{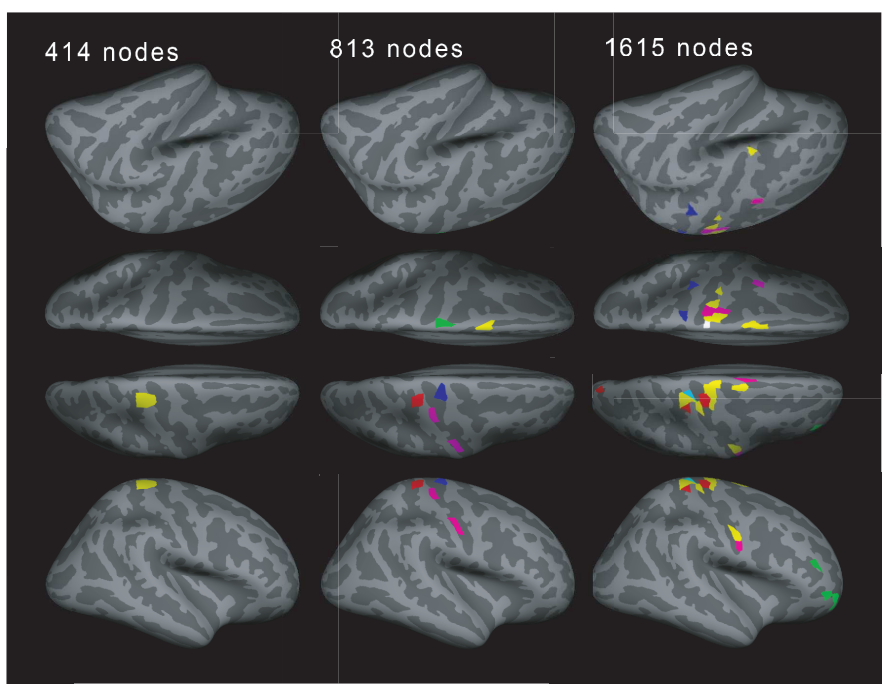}
 \caption{
 Example for the motif distribution in one subject (19 years old) for different spatial resolutions. 
 Different number of regions of interests (ROIs, left: 414, middle: 813, and right: 1615) with different views: left lateral view, left superior view, right superior view, and right lateral view in order from the top. yellow: motif 3, magenta: motif 4, cyan: motif 7, red: motif 8, green: motif 9, blue: motif 10 (motifs 1, 2, and 5 were not present). 
  }
 \label{no_motif_collated}
\end{figure}

How consistent are motifs across different resolutions? 
If a region splits into several regions for higher resolutions, do the daughter regions show the same motif as the parent region? 
Interestingly, this does not seem to be the case. 
Figure~\ref{no_motif_collated} shows a typical example of motif locations for one subject. For 414~nodes, one location shows motif~3. 
One part of this location then shows motif 8~for the resolution of 813~nodes. 
For 1615~nodes, part of the original region shows motif~3 again, whereas another part shows motif~9. 
This is an interesting point when talking about the connectivity between brain regions: A pattern shown at the low resolution level (e.g. cortical regions or Brodmann areas) may consist of several diverse patterns when observing connectivity at a higher resolution.


\section{Discussion} 

We observed how the distribution (fingerprint) of characteristic motifs of single network nodes changed for human structural connectivity (diffusion tensor imaging) during brain development.
We studied the role of both temporal scales (age groups from 12--23 years) and spatial scales (414, 813, and 1615~nodes) and found that, first, the number and diversity of motifs in a network are strongly correlated.
Second, comparing different scales, the number and diversity of motifs varied across the temporal (subject age) and spatial (network resolution) scale (with certain motifs only occurring at one spatial scale or for a certain age range).
Third, the sub-regions of a node, using a higher spatial resolution, may or may not include the original motif at the lower resolution.
Therefore, both the type and localisation of motifs differ for different spatial resolutions.
This indicates that spatial resolution is crucial when comparing characteristic node fingerprints given by topological and spatial network features.
This result is in line with previous studies that observed the role of sampling on brain network properties \citep{Antiqueira2010} and it additionally shows the influence of the temporal scale.

Networks can be characterized at different levels.
Aggregate measures that characterize a network as a whole can be used to distinguish different network types, e.g. small-world \citep{Watts1998, Sporns2000a, Hilgetag2000a}, scale-free \citep{Eguiluz2005, Kaiser2007}, modular \citep{Hilgetag2000a}, or hierarchical \citep{Kaiser2010editorial}.
Networks can also be characterized at the level of individual components.
Looking at the topological features of individual nodes, a first study by \citet{Passingham2002} has linked the network features of macaque cortical areas with the function of each area.
Whereas this approach considered all nodes of the network, there are also ways to search for 'special' nodes of a system.
Following the ideas of scale-free networks, brain areas with a large number of connections---so-called network hubs---might be crucial for integrating or distributing information \citep{Kaiser2007}.
\citet{Sporns2007} were able to detect and classify different kinds of hubs of cortical structural connectivity.
However, hubs are just one type of outliers where one or more node features (here: the number of connections of a node) differ from the average value of nodes in the network.
In two previous manuscripts \citep{Costa_2009, Echtermeyer2011}, we have developed a tool that systematically searches for and classifies nodes which differ from the majority of nodes in a network.
This approach includes the joint analysis of multiple features, which include topological as well as spatial properties \citep{Costa2007}.

Our measures are based on outliers, which do not show the overall trend as shown in previous studies (spatial resolution dependent changes: \citet{Zalesky_2010b, Bassett2010, Fornito2010, Hayasaka2010}, age dependent changes: \citet{Fan2010, Hagmann2010b, Fair2009, Uhlhaas2009}, Table~\ref{table_previous_studies}). 
Thus, it's difficult to directly relate our results to their findings. 
However, our report that edge density decreased with increased spatial resolution is consistent with previous studies of structural connectivity \citep{Zalesky_2010b, Bassett2010}.
For the temporal resolution, most previous studies showed no overlap with the age range of our study.
\citet{Hagmann2010b} performed a study with overlapping temporal scales, but only observed ages up to 18 years and thus only about half of the temporal scales considered in our study. 
Studies with larger overlap focussed on functional rather than structural connectivity \citep{Fan2010, Uhlhaas2009}.  

Here we noted that surface areas of used region-of-interest(ROI)s varied around 33\%--43\% compared to their mean (shown only for the cortical ROIs because subcortical ROIs were not subdivided, Table~\ref{table_surf_area_stat}). 
This variation stems from the variation of anatomical regions' surface areas (ratio of its standard deviation to its mean is 73.5\%).  
The ratio of interquartile range to median showed comparatively small ROI variation \citep{Fornito2010}.
To keep the topological consistency, this might be the best we can do. 
At least, this is better than using just anatomical regions whose variation is much higher (73.5\% variation). 

This node fingerprint study is a proof of principle for the current technique showing the influence of spatial and temporal scales on network comparison.
However, the current results could also be influenced by the following points.
First, the number of subjects per age group was not identical (Table~\ref{table_age_cat}).
Therefore, findings in age groups with fewer members might have been influenced by relatively few outliers.
Whereas an identical number of subjects per group is desirable, for our study there was also a trade-off with the age range: using the same subject numbers would have led to groups where the age ranges differ, e.g. covering 3~years for one group and 5~years for another.
We decided for a different number of subjects per group in order to perform the comparison across the temporal scale.
Second, single node motifs become more robust for larger network sizes.
As node motifs are outliers, only a small percentage of all nodes will be characterized as node motifs.
Therefore, for low-resolution networks, node motifs might only occur for few subjects.
In addition, their number might be a poor estimate of the underlying frequency in human connectivity networks.
For this reason, we have only reported networks with at least 414~nodes, leaving out networks with a parcellation into 110 cortical and subcortical regions.
Third, node features but not necessarily motifs depend on the edge density of the network.
In our developmental networks, we observed significant edge density changes for different spatial resolutions ranging from more than 7\% for 414~nodes to 2\% for 1615~nodes.
However, for each network resolution, edge densities remained comparable for different temporal scales (age groups) (Fig.~\ref{fig_edge_density}).
Fourth, deterministic tracking cannot capture crossing fibres which can be done in probabilistic tracking \citep{Behrens2007}, in Diffusion spectrum imaging \citep{Wedeen2008, Bassett2010}, or in high-angular resolution diffusion imaging (HARDI, \citet{Tuch2004, Zalesky_2010b}). 
However, fibre length distributions, influencing our three spatial features, were similar across resolutions (Fig.~\ref{fibre_length_dist}). 

Network Science has led to a wide range of tools for analysing neural systems \citep{Sporns2004, Costa2007a, Bullmore2009, Rubinov2010, Kaiser2011tutorial}.
Whereas the characterization of individual networks, e.g. as small-world or scale-free, is now possible, comparing different networks is still a challenge.
Our study shows how node properties, given by characteristic single-node motif fingerprints, can be compared {\em between} different networks.
Studying single node properties is only one way to characterize networks with alternatives being (a)~the study of global properties including local and global efficiency \citep{Latora2001,Achard2007} as well as modularity \citep{Newman2006a}, (b)~the comparison of the cluster organisation of different networks, and (c)~direct comparison of network matrices \citep{Crofts2009}. 
However, it is problematic for these measures that network features are influenced by the number of nodes and edges of a network; feature changes between networks might simply be due to different edge densities in the compared systems \citep{vanWijk2010}.
For our networks, edge densities did not significantly differ across age but they differed across spatial resolutions.
Whereas the underlying topological and, to some extent, spatial features might have changed in our networks, our motif fingerprints do not depend on the absolute values for network measures but on their distribution: an overall increase or decrease in the average value of a measure will have less of an influence on the number of node motifs as these depend on statistical outliers and not on absolute values of a measure.

Nonetheless, nodes that present characteristic node motifs, as well as hubs, are rare within networks.
These fingerprints can therefore only be applied for high-resolution networks ($\geq$414~nodes) and large groups of subjects.
Fortunately, large subject cohorts are currently being recruited in several initiatives including the Human Connectome Project for structural connectivity and the 1000 Functional Connectome Project \citep{Biswal2010}.
In addition, fingerprints will also be useful for high-resolution networks at the scale of the micro-connectome observing connections between individual neurons \citep{DeFelipe2010}.
Another potential application lies in the analysis of multi-electrode array recordings, which nowadays can record from more than 4,000~channels \citep{Sernagor2010}.
We therefore made this tool available within the CARMEN initiative (\url{http://www.carmen.org.uk}) for developing electrophysiology analysis tools \citep{Smith2007}.
The tool is also available on our website (\url{http://www.biological-networks.org/}).

\section{Conclusion} 
In this study we found that (a)~node motifs change over time and can therefore be used to find characteristic changes in the developing brain and (b)~motifs change over spatial scales, therefore comparisons between studies are only meaningful when the same (or a comparable) spatial resolution has been used.
Our results also indicate that spatial resolution has a higher effect on topological measures whereas spatial measures, based on fibre lengths, remain more comparable between resolutions. 
As node motifs are based on topological and spatial properties of brain connectivity networks, these conclusions are also relevant to other studies using network analysis.
Another important aspect is the analysis of differences between healthy controls and subjects with brain disorders that can arise during brain network development, such as schizophrenia and epilepsy; we hope that our node fingerprint approach will be useful for detection and specification in these cases.

\section*{Disclosure/Conflict of Interest} 
The authors declare that the research was conducted in the absence of any commercial or financial relationships that could be construed as a potential conflict of interest.

\section*{Acknowledgements} 
Marcus Kaiser and Christoph Echtermeyer were funded by EPSRC (EP/G03950X/1) and the CARMEN e-science Neuroinformatics project (\url{http://www.carmen.org.uk}) funded by EPSRC (EP/E002331/1).
Marcus Kaiser and Cheol Han also acknowledge support by the WCU program through the National Research Foundation of Korea funded by the Ministry of Education, Science and Technology (R32-10142). 

\bibliographystyle{abbrvnat} 
\bibliography{p_citations}

\FloatBarrier
\newpage
\section*{Tables}
\begin{table}[h]
  \caption{\label{table_age_cat}Ranges of age-categories.}
  \begin{center}
  \begin{tabular}{lll}
  \hline
  \textbf{category} & \textbf{age-range} & \textbf{number of subjects}\\
\hline
  1 & 12--14 years & 9\\
  2 & 15--17 years & 20\\
  3 & 18--20 years & 16\\
  4 & 21--23 years & 8\\
 \hline 
  \end{tabular}
  \end{center}
\end{table}
\begin{table}[h]
  \caption{\label{table_surf_area_stat}Region of Interest (ROI) surface area (mm\textsuperscript{2}) statistics with ratio of interquartile range to median. }
  \begin{center}
  \begin{tabular}{llll}
  \hline
  \textbf{parcellation} & \textbf{median} & \textbf{interquartile range} & \textbf{ratio}\\
\hline
aparc   &   1935     &   2811       &    1.45 \\
414     &    395     &    183       &    0.46 \\
813     &    195     &    101       &    0.52 \\
1615    &    95      &     56       &    0.59 \\
 \hline 
  \end{tabular}
  \end{center}
\end{table}
\begin{table}[h]
  \caption{\label{table_motif_properties}
    Properties of most frequent motifs. Symbols $\cdots\ $, $\uparrow$, and~$\downarrow$ indicate normal, elevated, and decreased values of normalised node degree~$K$,
normalised average degree~$r$,
coefficient of variation of neighbours' degrees~$cv$,
locality index~$loc$,
clustering coefficient~$cc$,
hierarchical clustering coefficient of level two~$cc_2$,
average connection length~$acl$,
maximum connection length~$mcl$,
average indirect reach~$air$, respectively.
}
  \begin{center}
  \begin{tabular}{llllllllll}
\hline
  \textbf{motif} 
& \textbf{$K$}
& \textbf{$r$}
& \textbf{$cv$}
& \textbf{$loc$}
& \textbf{$cc$}
& \textbf{$cc_2$}
& \textbf{$acl$}
& \textbf{$mcl$}
& \textbf{$air$} \\
\hline
  3 &$\downarrow$&$\downarrow$&$\downarrow$&$\uparrow$&$\uparrow$&$\downarrow$&$\cdots$&$\downarrow$&$\downarrow$ \\
  4 &$\cdots$&$\uparrow$&$\cdots$&$\uparrow$&$\downarrow$&$\cdots$&$\cdots$&$\downarrow$&$\cdots$ \\
  5 &$\downarrow$&$\cdots$&$\downarrow$&$\uparrow$&$\downarrow$&$\uparrow$&$\uparrow$&$\uparrow$&$\uparrow$ \\
  8 &$\downarrow$&$\downarrow$&$\uparrow$&$\uparrow$&$\downarrow$&$\uparrow$&$\uparrow$&$\ldots$&$\uparrow$ \\
  9 &$\downarrow$&$\cdots$&$\downarrow$&$\uparrow$&$\uparrow$&$\uparrow$&$\uparrow$&$\uparrow$&$\uparrow$ \\
  10 &$\downarrow$&$\cdots$&$\downarrow$&$\uparrow$&$\downarrow$&$\uparrow$&$\cdots$&$\downarrow$&$\cdots$ \\
\hline
  \end{tabular}
  \end{center}
\end{table}

\FloatBarrier
\newpage
\section*{Figures}
%
%

%
%
%
%

%
%
%
%

%
%
%
%

%
%
%
%

%
%
%
%

%
%
\FloatBarrier
\appendix
\section*{Appendix}
\renewcommand{\thetable}{S\arabic{table}}
\setcounter{table}{0}
\begin{table}[h]
  \caption{\label{table_previous_studies}
    Comparison of our findings with previous studies (SC:~structural connectivity, rsFC:~resting state functional connectivity, m/o:~months-old, y/o:~years-old, ED:~edge density, CC:~clustering coefficient, CPL:~characteristic path length, ADC:~apparent diffusion coefficient, FA:~fractional anisotropy, $\gamma$:~$\mbox{CC/CC}_{\mbox{rand}}$, $\lambda$:~$\mbox{CPL/CPL}_{\mbox{rand}}$). } 
  \begin{center}
  \begin{tabular}{|p{0.28\textwidth} | p{0.72\textwidth} |}
  \hline
  \textbf{study}  & \textbf{findings} \\
\hline
\textbf{Spatial scale:} & As the number of nodes increases, \\
Ours                    &  in SC (414, 813, and 1615 nodes), $\downarrow$~ED, $\uparrow w$ (See Fig.~\ref{fig_w_k_distribution}) \\
\citet{Zalesky_2010b}   &  in SC (6 different scales between 82 and 4000 nodes), $\downarrow$~ED, $\uparrow$~small worldness, $\uparrow \gamma$, $\downarrow$~global efficiency, changes in nodal rank degree and betweeness centrality \\
\citet{Bassett2010}     &  in SC (12 different scales between 54 to 880), $\downarrow$~ED, $\uparrow \gamma$, $\uparrow \lambda$, conserved hierarchy, $\uparrow$~Rentinan scaling\\
\citet{Hagmann2010b}    & in SC (66 and 241 nodes), $\downarrow$~ED, $\uparrow$~CC, $\downarrow$~Efficiency, $\downarrow$~Node strength \\
\citet{Fornito2010}     &   in rsFC (7 different scales between 84 and 4320 nodes), $\downarrow$~average correlation, $\uparrow$~size of largest component, $\downarrow$~path length, $\uparrow$~CC (for low ED), $\downarrow$~CC (for high ED) $\uparrow$~small worldness, changes in nodal rank degree \\
\citet{Hayasaka2010}    &   in rsFC (voxel-based nodes with 3 different voxel sizes and region-based nodes), changes in node degree distribution  $\uparrow$~CC, $\uparrow$~path length (while increasing voxel sizes, but it decreased with region based node) \\
\hline 
\textbf{Temporal scale:} & As the age increases, \\
Ours                     &  in SC (12 to 23 y/o), a characteristic pattern over spatial scales (See Fig.~\ref{fig_w_k_distribution}) \\ 
\citet{Hagmann2010b}     & in SC (18 m/o to 18 y/o), $\downarrow$~mean ADC, $\uparrow$~mean FA, $\downarrow$~ED, $\downarrow$~CC, $\uparrow$~Efficiency, $\uparrow$~Node strength, $\uparrow$~SC-FC correlation, no changes in modularity and no major changes in module composition after 2~y/o, no significant changes in betweeness-centrality \\
\citet{Fan2010}          &  in SC (1 m/o, 1 y/o, 2 y/o, and adult), $\uparrow$~global efficiency, $\uparrow$~cost efficiency, a peak at 2~y/o in local efficiency and modularity, $\uparrow$~size of largest component in modules, changes in module assignment and participation coefficient  \\
\citet{Fair2009}         &   in rsFC (7 to 31 y/o), no significant changes in optimized modularity Q, CC and CPL. Changed module assignment and the number of long-distance correlations~$\uparrow$. \\
\citet{Uhlhaas2009}      &   in EEG (6 to 21 y/o), strong correlation between neural synchrony and cognitive performance. As ages increase, the neural synchrony was increasing in early adolescence, then decreasing in late adolescence, and finally increasing again in adult \\
 \hline 
  \end{tabular}
  \end{center}
\end{table}

\end{document}